\documentstyle[aps,preprint]{revtex}
\begin{document}
\title{\bf Simple Mechanical Equivalents of Stepping Rotary 
Dynamics in F$_1$-ATPase
}
\author{
 A.V.~Zolotaryuk$^{1,2},$
V.N.~Ermakov$^{1,2}$,
P.L.~Christiansen$^2$, 
B.~Nord\'{e}n$^{3}$, and 
Y.~Zolotaryuk$^{1,2}$
}
\address{
$^{1}$Bogolyubov Institute for Theoretical Physics,
03143 Kyiv, Ukraine \\
$^{2}$Section of Mathematical Physics,
IMM, Technical University of Denmark, DK-2800 Lyngby, Denmark \\
$^{3}$Department of Physical Chemistry, Chalmers University of
Technology, S-412 96 Gothenburg, Sweden
 }
\date{\today}
%\wideabs{
\maketitle

\begin{abstract}

Two simple (rotator and one-particle) mechanistic models 
are suggested to describe simultaneously at a  minimal level
of sophistication two basic
functions of F$_1$-ATPase:  a motor regime 
 driven by ATP hydrolysis and its 
 inverted function as ATP synthesis. This description 
is consistent with the so-called rotary binding-change
mechanism, a milestone of functioning ATP synthase, and
uses  a stepping (driving) function associated with
two sequences of time instants, at which hydrolysis and
synthesis reactions occur. It is useful to analyse 
experimental data and numerical simulations indeed predict
 corresponding dynamic behavior.

\end{abstract}


%\twocolumn
%}

Recently, the modeling of molecular motors - enzymes which
transduce chemical energy into directed mechanical motion  -
based on the idea of gaining useful work by rectification
of zero-mean noise fluctuations has attracted
considerable attention \cite{a-r} and several models
exploiting the so-called {\it ratchet}
 mechanism have been elaborated
(see, e.g., Refs.~\cite{ab,p-a,dv,ma,cf,puk,lm,bko}, 
to mention a few).
In this context, ATP (adenosinetriphosphate) synthase
 shown schematically in Fiq.~1, being a realistic molecular engine
of great interest nowadays \cite{bba}, should also  
be studied from the point of view of biological 
physics \cite{fwa}. This machinery
is composed of two rotary motors: a membrane-embedded
unit F$_0$ and  water-surrounded F$_1$-ATPase
(also called F$_1$) connected by a coiled-coil
 $\gamma$ subunit.
ATP synthase works as a reversible motor-pump
machine: the proton flow through F$_0$ is believed to
create a clockwise
(when F$_1$ is viewed from  the F$_0$ side) torque that turns
$\gamma$, whereupon
  ATP molecules are sequentially synthesized at three 
catalytic sites, one on each $\beta$ subunit.
Vice versa, ATP hydrolysis, which also occurs sequentially
on $\beta$'s, but in the opposite direction, has been
demonstrated to make $\gamma$ rotate
 backwards, converting F$_0$ into a proton pump.
In this case, $\gamma$ is driven by sequential 
hinge-bending motions of $\beta$'s, like 
 a crankshaft by the pistons in a car engine (for more 
details see Ref.~\cite{wo}). In this paper,
 we focus on F$_1$-ATPase and consider its operation
in both hydrolysis (as a motor) and synthesis (as a
synthesizer) directions, using in parallel two simple 
mechanistic analogs: a plane rotator and a particle 
subjected to a periodic potential. Our description is
consistent with Boyer's binding-change mechanism \cite{bo,a-w},
the recent findings of Cherepanov {\it et al.} and 
Junge {\it et al.} \cite{cmj}
on elasticity properties of the $\gamma$ subunit,
as well as with the experimental results of 
Yasuda {\it et al.} \cite{y-y}.  
 
The structure of F$_1$-ATPase and the mechanics of motions
within it are too complex to allow a detailed
description  of all interactions and motions of its different 
parts, i.e., the three $\beta$ and one
$\gamma$  subunits \cite{bba,wo,bo,a-w,sej,n-k,k-y}.
Therefore, it would be useful to
 describe this very sophisticated 
three-dimensional system, using
 simple mechanical equivalents (springs, particles, etc.)
and keeping the main features
of rotary dynamics found in previous studies,
such as the modeling of  Oster and Wang \cite{wo}
and others \cite{bba,cmj,y-y,sej,n-k,k-y}.
This approach, ``from complexity to simplicity'',  is often 
used in biological physics. The typical example of such a
  modeling is the propagation of a
nerve impulse on the giant axon of the squid,
where insight in the dynamics of the original Hodgkin-Huxley
equations \cite{hh} has been obtained by reduction to the
minimal FitzHugh-Nagumo system \cite{fn}.

Here the three-dimensional and four-body interaction
in the $\beta_3\gamma$ subcomplex is suggested to be
 effectively described as a coupling of a planar rotator of length
$R_0$ centered in the middle of an equilateral triangle
(see left panels of Figs.~2 and 3) with three
equivalent catalytic sites (denote them by numbers 
1, 2, and 3) at the vertices
of the triangle. Obviously, this rotor-stator interaction,
 resulting in a crankshaft-like rotation of $\gamma$
(and cooperative catalysis on $\beta$'s), is
 a periodic function: $U(\theta +2\pi)=U(\theta)$, where
 $\theta$ denotes
the angular position of the rotator (positive if 
 counter-clockwise). The mechanical equivalent of a
  driving torque on $\gamma$ can be chosen as
  a stretched spring or an effective
particle displaced from equilibrium
 in the periodic potential $U(\theta)$ as illustrated in
the left and the right panels of Figs.~2 and 3, respectively.
 At each site $i=1,2,3$, a spring $K_i$ is attached
 connecting this site
with the rotator. All the three springs are supposed
to be identical, but only one spring is allowed to be
switched on, while the other two are switched off, at a time.
Then  sequential switching the springs will result in
a power stroke on $\gamma$.
Without loss of generality,
 the rotor-stator potential (given in units of the
 hydrolysis energy $W \simeq 80$ pN$\cdot$nm) can be
written as
\begin{equation}
U(\theta)= [r(\theta) -a]^2/ (l-a)^2 ,
\label{1}
\end{equation}
where $a=d/R_0$ is the length of each spring $K_i$
being undistorted, $r(\theta)= \sqrt{1+(1+a)^2-2(1+a)\cos
\theta}$ the instantaneous spring
length, and $l=r(\pm 2\pi/3)$. 

Let us now describe how our spring system operates
in both the hydrolysis and synthesis directions and
how it can be coordinated with ATP hydrolysis/synthesis
 reactions at the catalytic sites of the $\beta$ subunits.
According to Boyer's hypothesis \cite{bo}
supported by the structural studies by Walker and
coworkers \cite{a-w},
and recently by direct observation of Noji {\it et al}. \cite{n-k},
each site  1, 2, or 3
can be found at least in one of the three states: T
(ATP binding), E (empty), and D (ADP binding), at a time.
Structurally,  they are arranged as T, E, and D 
counter-clockwise (see Figs.~1-3) and can be put on the
$\theta$ axis at the lattice (``catalytic'') sites with
spacing $2\pi/3$, as shown in the right panels of 
Figs.~2 and 3. The dimensionless periodic (with period $2\pi$)
potential $U(\theta)$ 
[not necessary of the form (\ref{1})] is 
supposed to be ``rigidly tied'' to this lattice,
so that its minima are always found at sites with 
state T. We assume this potential to satisfy the normalization
conditions: $U$(T)$=0$ and $U$(D)$=1$. The last constarint
means that the potential energy of the effective particle
in state D is equal to the free hydrolysis energy $W$. 

Consider first  the hydrolysis direction when F$_1$
operates as a motor (see Fig.~2).
Let initially the $\gamma$ subunit (rotor) be found at equilibrium, 
performing there thermal fluctuations.
 This state is represented by
 spring $K_1$ being undistorted and switched on state T,
 while the springs at the other sites ($K_2$ and $K_3$)
 are switched off. In the particle equivalent, 
this situation is represented in the right
panel of Fig.~2(a) by a particle fluctuating in the vicinity
of one of the minima of the potential $U(\theta)$, i.e.,
in state T. When an ATP molecule
settles into site 2 which is found at this time in
 state E, it appears bound there,
resulting in ATP binding: E $\rightarrow$ T.
According to the rotary  binding-change mechanism \cite{bo,a-w},
this transition
 implies the two conformational changes at the next two sites: 
release of the inorganic  phosphate (P) and the ADP 
(adenosinediphosphate) molecule
 from site 2 (D $\rightarrow$ E), and
 hydrolysis of the ATP molecule located at site 3
 (T $\rightarrow$ D). After these state transitions have occured,
  spring $K_1$ is switched off, while spring $K_2$ is switched on,
causing the power stroke on the rotor as demonstrated by
Fig.~2(b). Therefore the rotor is driven forward before it
reaches a new equilibrium state, stepping forward
 by $2\pi/3$. Correspondingly, as shown in the right panel,
the potential $U(\theta)$ steps 
forward (to the right) by $2\pi/3$, so that the particle 
appears to be lifted uphill at the level $U=1$, thereafter 
sliding down and finally dwelling in the next potential 
minimum before the next sequence of conformational changes
takes place. However, if occasionally two sequential 
conformational transitions occur very close in time
resulting in two almost simultaneous potential steps forward,
the particle appears on the positive slope of the potential
$U(\theta)$, sliding thereafter down backwards. When the time
between two sequential transitions is still short, but long 
enough for the particle to make a descent close to equilibrium,
the double potential step forward will result in the double
sliding down on the negative slope. Indeed, both such 
occasional steps of $\gamma$ loaded by an actin filament
were observed in experiments \cite{y-y,n-k,k-y}. 
We denote the sequence of time instants
 when hydrolysis events occur by 
$\{ t_i^+\}_{i=1}^\infty~$ and call it a ``hydrolysis sequence''.

Consider now the case, when an external torque
(e.g., from the side of F$_0$)
is applied to the $\gamma$ subunit clockwise, as
 shown in the left panel of Fig.~3(a). According to
 Cherepanov {\it et al.} and Junge {\it et al.} \cite{cmj},
 the external energy of torsion
 is stored as an elastic strain energy of $\gamma$.
When the rotator turns clockwise totally by $2\pi/3$,
 the torsional energy
is liberated for the synthesis of one ATP molecule
 at the site being in state D. Then spring $K_1$ is switched off,
while spring $K_3$ is on, bringing the system to the zero-energy 
level ($U=0$). Equivalently, the particle ``drops down'' to
equilibrium state T [see Fig.~3(b)]. 
These mechanical equivalents are again in
accordance with Boyer \cite{bo} for ATP synthesis,
when the conformational change D $\rightarrow$ T implies
the transitions at the other two sites:
T $\rightarrow $ E and  E $\rightarrow $ D.
We denote the sequence of time instants when synthesis
events occur by $\{ t_j^-\}_{j=1}^\infty$ and call it
a ``synthesis sequence''.

Summarizing, one can present the diagram shown in Fig.~4,
which describes the elastic strain energy of
the torsional rotor-stator interaction as a function of
$\theta$, combining both the
hydrolysis and synthesis directions. 
Arrows show the rotational direction
under hydrolysis ($\theta >0$, {\it motor} regime)
and synthesis ($\theta <0$, {\it synthesizer} regime).

In experiments  \cite{y-y,n-k,k-y},
the $\gamma$ subunit was loaded by an actin 
filament rotating in a viscous solution.
Therefore, we approach the overdamped limit and
the equation of motion for the rotator 
 driven by a periodic potential $U(\theta)$
 fluctuating stepwise forward and
backwards by $2\pi/3$ (due to hydrolysis power strokes
and an external load torque ${\cal T}_l$) reads
\begin{equation}
\Gamma \dot{\theta}=- W \partial_{\theta}
U[\theta - 2\pi S(t)/3]- {\cal T}_l(\theta,t) + \xi(t) ,
\label{2}
\end{equation}
where $\Gamma$ is a viscous friction coefficient,
$U(\theta)$ is  normalized
by $U(2\pi n)=0$ and $U[2\pi(n \pm 1/3)]=1$,
 $n = 0, \pm 1, \ldots$~, and $\xi(t)$ is the Brownian torque
 with the auto-correlation
function $\langle \xi(t)\xi(t')\rangle =2\Gamma  
k_B T \delta(t-t')$,  where $k_B$ denotes
Boltzmann's constant and $T$ is the absolute temperature.
The stepping (driving) function
 \begin{equation}
S(t)= \sum_{i=1}^\infty\Theta(t-t^+_i)
- \sum_{j=1}^\infty\Theta(t-t^-_j) ,
\label{3}
\end{equation}
where $\Theta(t) = 0$ for $t<0$ and
 $\Theta(t) = 1$ for $t \ge 0$,
is defined  through the two time sequences
$\{ t_i^{+}\}_{i=1}^\infty$  and
$\{ t_j^{-}\}_{j=1}^\infty$ indicating the time instants
when hydrolysis and synthesis reactions occur,
respectively.

The hydrolysis sequence $\{ t_i^+\}$ is a random
process determined by ATP concentration ([ATP]).
We define it through two characteristic duration
times $\Delta t_{\mbox{\tiny D}}$
and $\Delta t_{\mbox{\tiny T}}$ by the recurrence
relation
\begin{equation}
t^+_{i+1} =t^+_i +\Delta t_{\mbox{\tiny D}} +2\zeta_{i+1}
\Delta t_{\mbox{\tiny T}},~i=1, 2, \ldots ,
\label{4}
\end{equation}
where each $\zeta_i \in [0,1]$ is a random value  with
 uniform distribution (since ATP concentration is supposed
to be constant in the solution).
The interval $\Delta t_{\mbox{\tiny T}}$ 
 is large for low [ATP],
but tends to zero as [ATP] is sufficiently
high. Therefore one can assume that
$ \Delta t_{\mbox{\tiny T}} = A_{\mbox{\tiny T}}/[\mbox{ATP}]$
with some constant $A_{\mbox{\tiny T}} > 0$.
The interval $\Delta t_{\mbox{\tiny D}}$
triggers the hydrolysis reactions
 by release of P and ADP on the next site of the stator.
In the limit of low concentration of
nucleotides (ATP and ADP), this interval is short, whereas
for high concentration,
the release of the hydrolysis products
 is impeded  and therefore some saturation for
 $\Delta t_{\mbox{\tiny D}}$ takes place.
As a result, one can assume that
$ \Delta t_{\mbox{\tiny D}} = \Delta t_{\mbox{\footnotesize st}}
 [\mbox{ATP}]/( C_{\mbox{\tiny D}} + [\mbox{ATP}])$
 with some constant  $C_{\mbox{\tiny D}} > 0$ and 
$\Delta t_{\mbox{\footnotesize st}}$ being
 the duration time of one step. At zero temperature,
it follows from the overdamped dynamics
governed by Eq.~(\ref{2}) that
  $\Delta t_{\mbox{\footnotesize st}} \rightarrow \infty$
  (an overdamped particle
approaches a potential minimum after an infinitely long time).
 However, at nonzero temperature, the time for
 one step becomes  {\it finite}, because in the vicinity
 of equilibrium,
the particle is ``captured'' by thermal fluctuations.
Thus, in the limit $a \rightarrow 0$,
the potential (\ref{1}) is reduced to the simple form
 $U=2(1-\cos\theta)/3$, admitting an explicit solution of
 Eq.~(\ref{2}) for each step if $\xi(t) \equiv 0$.
 Since the average amplitude of thermal
 fluctuations is $\sqrt{3{\cal D}/2}$, one finds
that $\Delta t_{\mbox{\footnotesize st}} \simeq
(3t_0/2)\ln\left(\sqrt{3}\mbox{cot}\sqrt{3{\cal D}/8}\right)$,
where $t_0=\Gamma/W$ is the time unit and
 ${\cal D}=k_B T/W$  the dimensionless strength of
white noise in Eq.~(\ref{2}). Since,
at room temperature $ k_BT \simeq 4$ pN$\cdot$nm,
we thus have ${\cal D} \simeq 0.05$, so that
$\Delta t_{\mbox{\footnotesize st}}$ can be estimated
for each viscous load $\Gamma$.

Averaging Eq.~(\ref{4}), one finds that
 $\langle t^+_{j+1}-t^+_j \rangle
=\Delta t_{\mbox{\tiny D}} +
  \Delta t_{\mbox{\tiny T}} $ and therefore the
rotational rate of $\gamma$ (number of revolutions per second) is
$ V = \left[ 3\left(\Delta t_{\mbox{\tiny D}} +
\Delta t_{\mbox{\tiny T}}\right)
\right]^{-1}$. Inserting here the dependences
of $\Delta t_{\mbox{\tiny D,T}}$ on [ATP] given above, 
one finds
\begin{equation}
 V =  { V_{\mbox{\footnotesize max}} [\mbox{ATP}] \over
 K_{\mbox{\tiny M}}
  +[\mbox{ATP}]^2/(C_{\mbox{\tiny D}} +[\mbox{ATP}]) }~,
\label{5}
\end{equation}
where $ V_{\mbox{\footnotesize max}} =
(3\Delta t_{\mbox{\footnotesize st}})^{-1}=
\left(\Delta t_{\mbox{\tiny D}}\right)^{-1}$
is a maximal average velocity (as
$[\mbox{ATP}] \rightarrow \infty$) and the constant
$K_{\mbox{\tiny M}} = A_{\mbox{\tiny T}}/
\Delta t_{\mbox{\footnotesize st}}$ can be identified as
 the Michaelis constant, because Eq.~(\ref{5}) is reduced
to the  Michaelis-Menten law in the limit
 $C_{\mbox{\tiny D}} \rightarrow 0$.
According to Yasuda {\it et al.} \cite{y-y},
 $K_{\mbox{\tiny M}} =0.8~\mu$M and $V_{\mbox{\footnotesize max}}=
4$ s$^{-1}$, and for these values
the dependence (\ref{5}) is shown in Fig.~5
(including the experimental data), with
 monotonic behavior for  $ C_{\mbox{\tiny D}}
 \le K_{\mbox{\tiny M}}$.
Similarly \cite{y-y}, one can assume that the one-step
duration depends on the 
 length of the filament $L$ as
$\Delta t_{\mbox{\footnotesize st}} \simeq g_0 + g_1 L^3$, and
substituting this expression into Eq.~(\ref{5}),
one finds for small  $ C_{\mbox{\tiny D}}$
 the dependence
$ V(L) \simeq \frac{1}{3}\left(A_{\mbox{\tiny T}}/[\mbox{ATP}]
 +g_0 + g_1 L^3 \right)^{-1}$ plotted in Fig.~6, 
where the constants 
$ A_{\mbox{\tiny T}}$ and  $g_{0,1}$ are fitted 
 to the experiments \cite{y-y}.

As described above, for a given random hydrolysis 
sequence (\ref{4}),
 each synthesis instant $t_j^-$ is defined as the time when
 the rotor, being at some time in
 state T, rotates backwards by $2\pi /3$.
Direct simulations of the dimensionless
($\tau = t/t_0$) equation (\ref{2}) with
the potential (\ref{1}) are presented in Fig.~7. 
Here the intervals $\Delta \tau_{\mbox{\tiny D,T}} = 
\Delta t_{\mbox{\tiny D,T}}/t_0$ in the sequence (\ref{4})
are given through the constants  $ C_{\mbox{\tiny D}}$ and 
$ K_{\mbox{\tiny M}}$, as well as the parameter 
$\Delta \tau_{\mbox{\footnotesize st}} =
\Delta t_{\mbox{\footnotesize st}}/t_0 
= W/3\Gamma V_{\mbox{\footnotesize max}}$
to be evaluated from experiments. Thus, using that
$\Gamma(L=1~\mu\mbox{m})
\simeq 1$ pN$\cdot$nm$\cdot$s and $V_{\mbox{\footnotesize max}}=
4$ s$^{-1}$ \cite{y-y}, one finds
$\Delta \tau_{\mbox{\footnotesize st}} \simeq 6.75.$
 In the case without load (curves 1 and 2), the average
velocity $\langle \theta \rangle /2\pi \tau
= \Gamma V/W$
is in good agreement with the direct observations (see Fig.~3 of
Ref.~\cite{y-y}) and the law plotted in Fig.~5 at 
  $ C_{\mbox{\tiny D}} = K_{\mbox{\tiny M}}$, where the
two velocities shown with dotted lines 1 and 2 correspond to
curves 1 and 2 in Fig.~7, respectively. Note that the
 constant  $ C_{\mbox{\tiny D}}$ in Eq.~(\ref{5}) 
implicitly describes the main feature of trajectories
at low [ATP] (see curve 2): 
on average the number of multi-steps exceeds 
that of steps backwards, as observed experimentally
\cite{n-k,y-y}. Next, when a load torque 
${\cal T}_l > 0$ is applied, the rotational rate
decreases with increase of this torque, as
illustrated by curve 3 in Fig.~7. Moreover, when the load
exceeds some threshold value, the motor operates in inverse,
as shown by curves 4 and 5.
 Similarly to Lattanzi and Maritan \cite{lm}, the
law (\ref{5}) can be modified by
subtracting a positive constant that controls the direction
of rotation if ${\cal T}_l >0$.

Thus, we have developed two (rotator and one-particle) 
physical models of archetypal simplicity, which
are consistent with the rotary binding-change 
mechanism \cite{bo,a-w} and the elasticity properties 
of the $\gamma$ subunit \cite{cmj}.
The cooperative rotational catalysis at the three
$\beta$ subunits is described through two 
 time sequences, each for switching hydrolysis and synthesis
reactions, by adjusting the statistics of switching
to satisfy the recent experimental results \cite{n-k,y-y}.
The models described in this paper are generic and simple;
 they do not depend on details of the periodic
potential $U(\theta)$. 
In the hydrolysis (motor) direction, the $\gamma$ subunit 
works in a ``passive'' regime; only all the $\beta$ subunits
are coordinated in the cooperative rotational catalysis.
In the reversible (synthesis) direction, $\gamma$ is ``active'',
causing the corresponding (again cooperative, but in
the inverse sequence) conformational
changes after its strain energy of torsion reaches the free
ATP hydrolysis energy. These important features are consistent
with both Boyer's binding-change mechanism \cite{bo,a-w}
 and the findings of Cherepanov {\it et al.} and Junge {\it et
al.} \cite{cmj}. The dependences of the model 
parameters on ATP concentration are general and 
physically motivated. In the framework of our description, 
the load torque ${\cal T}_l(\theta, t)$ generated by
the F$_0$ part can further be involved explicitly 
resulting in a general motor/pump model of ATP synthase. 

   We also conclude that the puzzle, how does the
binding-change mechanism work, may be essential not only
for understanding the chemistry (dissipative catalysis) of
creation of ATP, one of the most important processes in life,
but also constitute a key physical problem behind the function
of molecular motors, such as design of man-made molecular
 devices.

We acknowledge partial financial support
from the European Union under the INTAS
Grant No.~97-0368 and the LOCNET Project No.~HPRN-CT-1999-00163.
We thank A.C.~Scott for stimulating and helpful discussions.

\newpage

\begin{center}
FIGURE CAPTIONS
\end{center}

FIG.~1. Schematics of ATP synthase adapted
from Ref.~\cite{k-y}. An asymmetric $\gamma$ shaft
rotates relatively to the hexamer formed by $\alpha$ and
three $\beta$ subunits arranged alternatively. The other
subunits, which constitute F$_0$ including the ``anchor''
part are not shown. The positive direction of $\gamma$
rotation and the directions of proton flow and rotational 
catalysis (sequential synthesis/hydrolysis reactions in the
hexamer) are shown by the arrows.

FIG.~2.
Spring (left panels) and particle (right panels) equivalents
of the rotor-stator system evolving in hydrolysis
direction. (a) Rotor (left) and particle (right)
are found in equilibrium (in state T).
(b) Power stroke caused by stretched spring $K_2$ will
rotate $\gamma$ counter-clockwise
(left). Sliding the particle down on the negative slope
of periodic potential $U(\theta)$, after it has moved  
forward by $2\pi/3$ (right).

FIG.~3.
Spring (left panels) and particle (right panels) equivalents
of the rotor-stator system evolving in
 synthesis direction.
(a) External torque drives the rotator clockwise
and elastic strain energy of the system is stored in spring $K_1$
(left). Lifting particle uphill in potential $U(\theta)$
(right).
(b) Release of elastic strain energy after synthesis takes place
(left). After the particle energy has reached the value 
$U=1$,  potential $U(\theta)$ steps
backwards by $2\pi/3$, allowing the particle to ``drop downhill''
to zero energy level (right).

FIG.~4.
Strain energy of the rotor-stator system against
 angular position $\theta$ in  hydrolysis
($\theta >0$) and synthesis ($\theta <0$).

FIG.~5.
Rotational rate of $\gamma$ against ATP concentration
calculated for three values of $C_{\mbox{\tiny D}}$,
using Eq.~(\ref{5}). Experimental results \cite{y-y}
shown by circles are given for comparison.

FIG.~6.
Rotational rate of $\gamma$ against the length of actin
filament: experimental results
(circles, squares, and triangles) and
 dependences $V(L)$ calculated
for three values of ATP concentration.

FIG.~7.
Typical trajectories for different regimes of F$_1$.
Curves 1 and 2 demonstrate pure motor regime (without load)
  for high (curve 1; $\Delta \tau_{\mbox{\tiny D}}
  =6.75$ and $\Delta \tau_{\mbox{\tiny T}} =0$)
  and low (curve 2; $\Delta\tau_{\mbox{\tiny D}} = 2$,
   $\Delta\tau_{\mbox{\tiny T}} =10$, [ATP] = 0.4 $\mu$M,
and  $C_{\mbox{\tiny D}}= K_{\mbox{\tiny M}}=0.8$ $\mu$M)
ATP concentration. 
The other trajectories illustrate mixed motor/synthesizer
regimes for different constant loads:
below threshold
 (curve 3; $\Delta \tau_{\mbox{\tiny D}} =1.7$,
 $\Delta \tau_{\mbox{\tiny T}} =20$, and
 ${\cal T}_l =20$ pN$\cdot$nm), nearby threshold
  (curve 4; $\Delta \tau_{\mbox{\tiny D}} =6.75$,
  $\Delta \tau_{\mbox{\tiny T}} =0$, and
   ${\cal T}_l  = {\cal T}_{\mbox{\footnotesize th}}
  \simeq 38.4$ pN$\cdot$nm), and above threshold
 (curve 5; $\Delta \tau_{\mbox{\tiny D}} =6.75$,
 $\Delta \tau_{\mbox{\tiny T}} =0$, and ${\cal T}_l =0.48$
 pN$\cdot$nm).

\end{document}